# Shading and Smothering of Gamma Ray Bursts

David Eichler[1] & Amir Levinson[2]

## ABSTRACT

The gamma ray burst (GRB) 980425 is distinctive in that it seems to be associated with supernova (SN) 1998bw, has no X-ray afterglow, and has a single peak light curve and a soft spectrum. The supernova is itself unusual in that its expansion velocity exceeds c/6. We suggest that many of these features can be accounted for with the hypothesis that we observe the GRB along a penumbral line of sight that contains mainly photons that have scattered off ejected baryons. The hypothesis suggests a baryon poor jet (BPJ) existing within a baryon rich outflow. The sharp distinction can be attributed to whether or not the magnetic field lines thread an event horizon. Such a configuration suggests that there will be some non-thermal acceleration of pick-up ex-neutrons within the BPJ. This scenario might produce observable spallation products and neutrinos.

The problem of baryon contamination of gamma ray bursts is well known. In order to be detectable at cosmological distances, a GRB must put out enough power $10^{50}$ erg s$^{-1}$ to drive away at least $10^{-3} M_\odot$ of baryonic matter over the duration of the burst, and this would be enough to obscure gamma rays originating within $10^{14}$ cm of the point of energy release. Because the maximum Lorentz factor $\Gamma$ of the bulk expansion could not be large compared to unity, however, the gammasphere would have to be placed well within this radius if the observed burst duration is to be only seconds, leaving the non-thermal nature of the spectra and short timescale a puzzle.

Suggested solutions for this problem include hypotheses that a) the GRB is powered by a merger of strange stars (Haensel, Paczynski, and Amsterdamski, 1992; Usov, 1998) which hold their baryons using strong interactions and b) the burst originates on field lines that thread an event horizon (Levinson and Eichler, 1993; Iwamoto, 1998; Paczynski 1998). Other proposals (e.g. Eichler et. al. 1989; Mezaros and Rees, 1992; Mochkovitch, R. et. al. 1993) involve some combination of geometry and centrifugal force that keep the axis of an accretion disk free of baryons, but they would need to be quantified and are harder to analyze.

If the baryon purity is enforced by an event horizon that forms during the same process that powers the GRB, then a baryonic outflow would be driven from the matter that had not yet fallen into the event horizon. It would most likely surround the BPJ that emerges from the event horizon and its photosphere would be considerably larger than that of the BPJ. The external observer could, therefore, see the primary emission from the BPJ only if looking down the hole it makes in the baryonic outflow. If too far off to the side, the observer might not see anything. A third possibility, however, is that the observer is offset from the BPJ axis enough that the central engine of the GRB is obscured, but the walls of the baryonic outflow that interface the BPJ are partly visible to the observer (see figure 1). An observer viewing from such a "penumbral" line of sight could see photons that were scattered from the walls that envelop the BPJ but not photons in its primary beam.

---

[1]Physics Department, Ben-Gurion University, Beer-Sheva 84105, Israel

[2]School of Physics & Astronomy, Tel Aviv University, Tel Aviv 69978, Israel; Levinson@wise.tau.ac.il

GRB 80425 has a high probability of being associated with the unusual supernova (SN) 1998bw (Galama et. al., 1998). Kulkarni, et. al. (1998) have reported radio emission beginning several days after the supernova, and argued that the GRB, being intrinsically weak if indeed associated with the SN, is then a member of a separate class of GRB's. Iwamoto (1998) has proposed a model in which the GRB is generated by an accreting black hole that was made during the same collapse that generated the SN. In his model, the GRB energy and associated afterglows, are intrinsically weak as argued by Kulkarni et. al.

We propose that GRB 80425 is in fact being viewed along a "penumbral line of sight" as defined above, that it represents a class of viewing angles along which GRB appear much weaker, and that it is perhaps the tip of an even larger class of buried GRB. There are several motivations for this given below, apart from the possible association of the GRB with a supernova. We envision that most of the gamma-ray emission in the BPJ is accompanied by a baryonic outflow. The latter may be a pre-existing star or it may be a wind driven from a neutron star as it merges. If it is a wind, we assume that the outflow velocity exceeds the BPJ photosphere divided by the burst duration, so that during the rise of the primary flux the baryonic outflow extends out to a radius much larger than the radius of the primary emission zone. Viewing the source from outside the primary beam we only see the scattered component. Failure to see X-ray afterglow could be attributed to our being outside the beaming cone of afterglow emission.

The variation of the scattered emission is anticipated to occur on a timescale of order the light travel time across the system (30 light seconds in the case of GRB 980425), provided the primary burst persists for a shorter time. Temporal substructure in the primary emission would be smeared out in the scattered component, rendering the time profile singly peaked. Having been scattered at least once, the spectrum would cut off sharply above several hundred KeV. The scattered component, though extending over a broader angular region, could be considerably weaker, depending on the detailed assumptions regarding the geometry and kinematics: The inner wall of the baryonic outflow may catch only a small fraction of the BPJ's emission, and the angle of incidence between photon and wall may be sharper than the angle between the wall and the observer, leading to a net loss of energy even in the Thomson scattering limit. Compton recoil in any case sets in at several hundred KeV.

The explanation of a single peak light curve is of course presented after the observation. Situations that may lead to different temporal characteristics (e.g. flash-in-the-pan effects) could be envisaged, but a prolonged, smeared light curve seems like a reasonable signature of scattering off baryonic matter that accompanies the burst. Softening of the spectrum where Klein Nishina effects set in would seem to be an unavoidable consequence of scattering.

As an illustrative example we consider a cylindrical wall illuminated by a beamed, point source located on the axis of the cylinder a distance $z_{ph}$ from the photosphere of the baryonic outflow. The cone of emission is taken to be coaxial with the cylinder with an opening angle $\eta_{max}$. The wall is assumed to move at a constant velocity, $\beta$, and to have an infinitely large Thomson depth. To simplify the analysis, the scattering cross section is assumed to be isotropic in the frame of the wall, and KN effects are ignored. Let us denote by $\mathcal{L}_{GRB}(\epsilon)$ the spectral energy distribution of the illuminating source. The power per unit energy per unit solid angle emitted in a direction $\hat{\Omega}$ by the section of the wall exposed to the observer is then given by

$$\mathcal{L}_{em}(\epsilon_s, \hat{\Omega}) = \frac{\sin\theta}{4\pi^2(1-\beta\cos\theta)^3} \int_{\eta_\theta}^{\eta_{ph}} \left\{ F(\eta)\sin\eta(1-\beta\cos\eta)^3 \mathcal{L}_{GRB}\left[\frac{(1-\beta\cos\theta)}{(1-\beta\cos\eta)}\epsilon_s\right] \right\} d\eta \qquad (1)$$

where $\theta$ is the angle between $\hat{\Omega}$ and the wall axis, $\eta_{ph} = \cot^{-1}(z_{ph/R})$, with $R$ being the cross-sectional radius of the wall, $\cot\eta_\theta = \max\{\cot\eta_{ph} - 2\cot\theta; \cot\eta_{max}\}$, and $F(\eta) = [4 - \tan^2\theta(\cot\eta_{ph} - \cot\eta)^2]^{1/2}$.

Note that when $\theta = \pi/2$ the lower limit $\eta_\theta = \eta_{ph}$ and the integral in eq. (1) vanishes, as it should. For a power law spectrum, viz., $\mathcal{L}_{GRB}(\epsilon) = \mathcal{L}_o \epsilon^{-p}$; $\epsilon_{min} < \epsilon < \epsilon_{max}$, eq. (1) reduces to

$$\mathcal{L}_{em}(\epsilon_s, \hat{\Omega}) = \frac{\mathcal{L}_o \epsilon_s^{-p}}{4\pi^2} G(\beta, \theta, \epsilon_s) \tag{2}$$

with

$$G(\beta, \theta, \epsilon_s) = \frac{\sin\theta}{(1-\beta\cos\theta)^{3+p}} \int_{\eta_\theta}^{\eta_{\epsilon_s}} \left\{ F(\eta) \sin\eta (1-\beta\cos\eta)^{3+p} \right\} d\eta, \tag{3}$$

where $\cos\eta_{\epsilon_s} = \min\{\cos\eta_{ph}; [1-(\epsilon_s/\epsilon_{max})(1-\beta\cos\theta)]/\beta\}$. As seen, the spectrum emitted by the wall is also a power law with an upper cutoff at energy $\epsilon_b = \epsilon_{max} \frac{1-\beta\cos\eta_{ph}}{1-\beta\cos\theta}$. Plots of $G$ versus $\beta$ for energies below the cutoff, and for different viewing angles, are exhibited in fig. 2.

The radio afterglow appears to come from an emission region moving with a bulk Lorentz factor of about 2 (Kulkarni et. al. 1998), (but see Loeb and Waxman, 1998). This could be because the jet has slowed down from higher values of $\Gamma$ by the time the radio afterglow is produced. But it could also be that the radio emission is generated by the forward shock of the ejected baryons. The required Lorentz factor of 2 or so implies a velocity only 3 or 4 times faster than the ejecta and could be accounted for by the acceleration of a shock front as it moves down the density gradient in the outer part of the presupernova star. The enormous ejecta velocity, five or six times that of a typical type 1C supernova, could be correlated with the appearance of gamma rays. One could argue that a low baryon density jet is formed in a much larger, less peculiar class of supernova, and that the gamma rays are usually smothered. Although not confident about the following point, we do not rule out the possibility that the baryonic output of SN 1998bw is low because it came directly off a preexisting neutron star that underwent a merger with some other object. This might account for the unusually high level of radioactivity in the ejecta as well as the high velocity.

The above hypothesis may also have the following observational consequences:

i) Shaded bursts, being effectively weaker in the direction of the observer, would have a different $V/V_{max}$ distribution from the usual ones, possibly even approaching an average value of 1/2. This is a retrodiction, since it has already been noted by Tavani (1998) that long duration, soft GRB's have a different distribution from the majority, which show a cosmological imprint in their distribution.

ii) The scattered hard X-/soft gamma rays should be linearly polarized, and it may be technologically possible to measure their polarization in the not too distant future.

iii) The coexistence of a medium that is low enough density to admit shock acceleration with dense baryonic material that envelops it could be the site of Be and B production in the Galaxy. Here we have assumed that buried gamma ray bursts are not uncommon. (In this regard , it is worth distinguishing between allowing gamma rays to escape, which requires extreme baryon purity on the scales of GRB photospheres, and allowing shock acceleration, which may be less demanding.)

iv) Particle acceleration within a very dense environment also implies neutrino production via pions if the accelerated particles are baryons. The possibility of high energy neutrino production in GRB has been considered by a number of authors.

With a nominal GRB rate of $10^{-6}$/galaxy-yr, process (iii) is irrelevant and and process (iv) is probably not observable with neutrino detectors now under construction (the largest being AMANDA) unless the total GRB output greatly exceed $10^{-4}$ erg/cm$^2$ (Eichler, 1994). However, if we are to consider the possiblity that any stellar collapse that yields a black hole also (much of the time, at least) yields a smothered GRB, then one could postulate that they are far more frequent than GRB, and perhaps almost as frequent as supernovae.

Consider now a BPJ emanating from a black hole that has formed in the midst of a supernova (or hypernova). Some neutron rich material should be expelled behind the rest of the ejecta by the same arguments that exist for GRB. Neutrons can cross field lines and drift into the BPJ. If they decay there, and the jet is already relativistic, then they immediately become cosmic rays in the frame of the jet. Moreover, they become the principal injection component to any shock acceleration that occurs in the jet. Estimating the infusion velocity v of neutron into the BPJ by equating the transverse momentum imparted to a proton at the wall to a significant fraction of the protons outward momentum $m_p u$ we obtain, neglecting the proton - neutron mass difference,

$$v^2 \sigma n = u/\delta t \qquad (4)$$

where n is the density of neutrons at the wall of the BPJ, $\delta t$ is the hydrodynamical timescale, and $\sigma$ is the neutron proton scattering cross section, whence

$$v/u \sim R/(N\sigma)^{1/2} \qquad (5)$$

where we have left aside dimensionless geometric factors. Here $N \sim nuR^2\delta t$ is the total number of neutrons, maybe $10^{54}$, and $\sigma$ is of order 5 barns, The quantity R is the radius at which the crossing takes place, which we take to be $10^{12}$ cm, so that the neutrons have not quite decayed out of the BPJ, but soon will decay inside it given that v is small compared to u. With these numbers, $10^{50.5}$ neutrons can be picked up by the BPJ. This would be enough to give of order $4 \times 10^{48}$ Beryllium nuclei (the needed quantity as estimated by several authors, e.g., Drury and Parizot, [1999] and references therein) if a fair fraction of these ex-neutrons make their way into CNO-rich ejecta. The detectability of the neutrinos, which would be made in comparable numbers to the number of pick-up neutrons but at a factor of 10 or so less energy, depends on the assumed bulk Lorentz factor in BPJ at the point of pick-up. Assuming the bulk Lorentz factor of $10^3$ and, say $10^{50}$ pick-up neutrons, gives of the order of $10^{50}$ 100 GeV neutrinos. This would be detectable by AMANDA at distances of an Mpc or so. Shock acceleration of the ex-neutrons could raise the average energy (and hence the detectability) of the neutrinos.

The various scenarios outlined above, if ever confirmed, would represent a qualitative manifestation of the hypothesized event horizon, which is responsible for the sharp transverse density and velocity gradients within the explosion. One should note, however, that any baryon-retaining mechanism, such as magnetic field lines threading strange matter, could also maintain sharp gradients provided that it is immersed in a baryonic outflow.

AL acknowledges support by Alon Fellowship and a grant from the Israeli Science Foundation. DE acknowledges a grant from the Israeli Science foundation. We are grateful to L. Drury, E. Waxman, and V. Usov for helpful conversations.

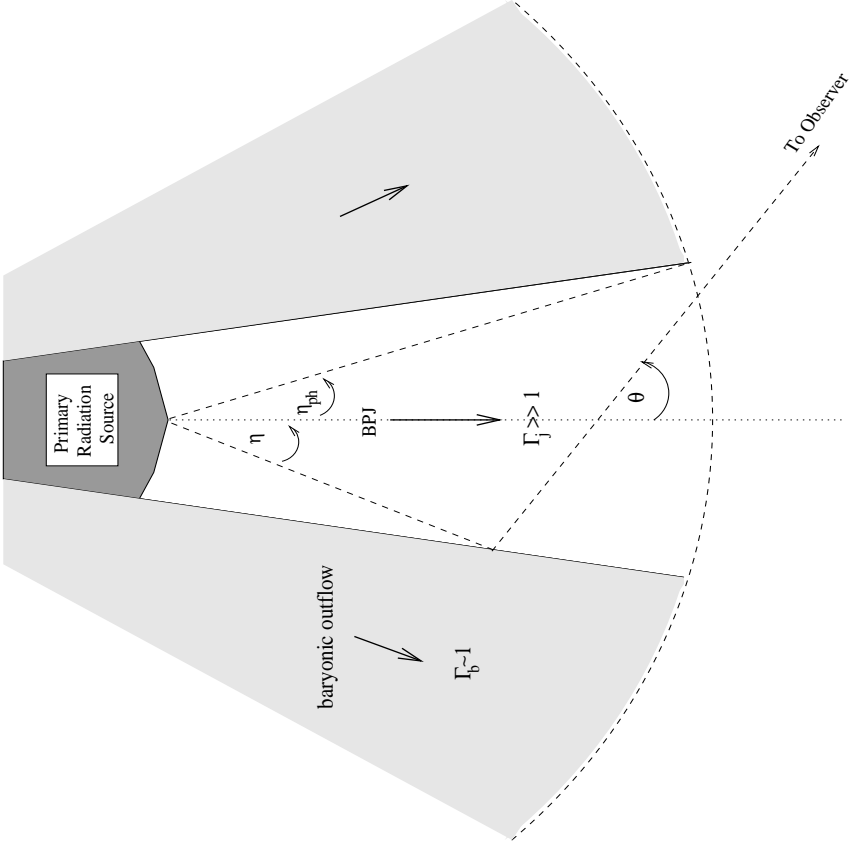

Fig. 1.— Schematic representation of the model. A baryon-poor jet (BPJ) is expelled from the central engine with a high Lorentz factor $\Gamma_j$. The BPJ is surrounded by a slower ($\Gamma_b$ of order unity), Thomson thick, baryon reach outflow. Gamma-rays emitted by the primary radiation source at angles $\eta > \eta_{ph}$ with respect to the BPJ axis are Compton scattered by the moving walls surrounding the BPJ. An observer viewing the source at an angle $\theta > \eta_{ph}$ cannot see the primary beam, but can see the flux scattered in his direction by the wall section exposed to him.

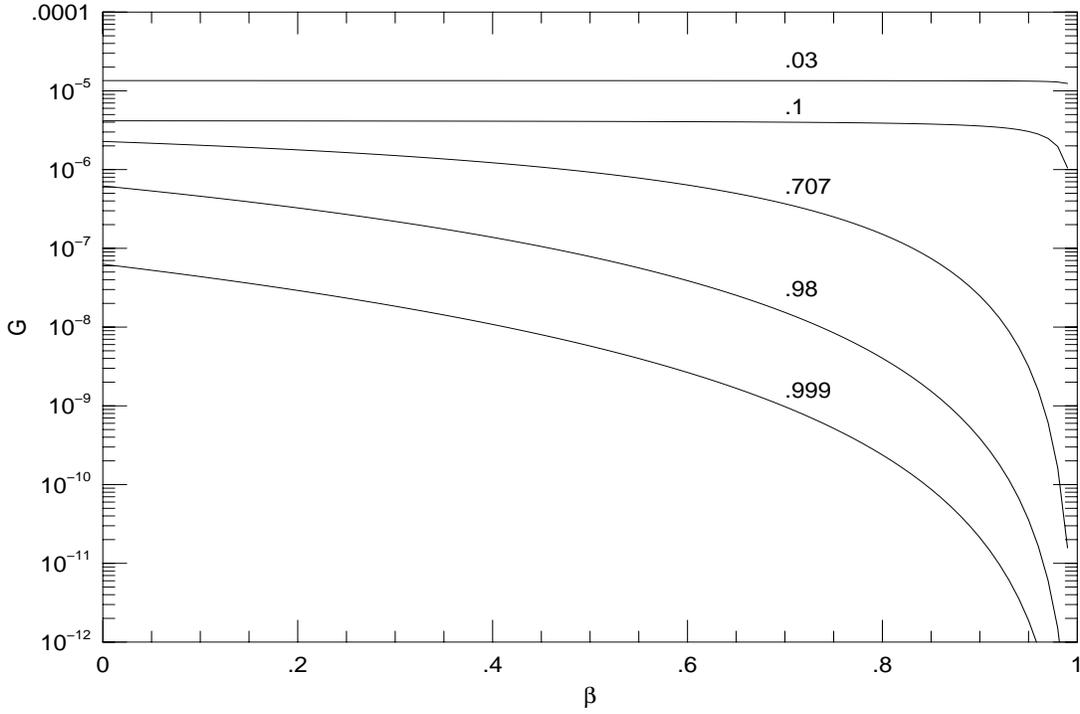

Fig. 2.— Plots of $G(\beta,\theta)$ (see eq. 3) versus $\beta$, for $p = 0.5$, $\sin\eta_{max} = 0.1$, and different viewing angles. The numbers that label the curves are the values of $\sin\theta$.